\documentclass[english,thm-restate]{lipics-v2021}
\usepackage[toc,page]{appendix}

\newcommand{\SoCGV}[1]{}
\newcommand{\ARXIV}[1]{#1}

\hideLIPIcs
\nolinenumbers 

\usepackage{graphicx} %package to manage images
\graphicspath{ {images/} }

\usepackage[scr=rsfs]{mathalpha}

\usepackage{subcaption}
\usepackage{booktabs}
\usepackage[ruled,vlined]{algorithm2e}
\usepackage{hyperref}
\usepackage{subfiles}
\usepackage{array}
\usepackage{bm}

\usepackage{amsfonts}
\usepackage{amsmath}
\usepackage{amsthm}

\usepackage{cleveref}

\DeclareMathOperator{\poly}{poly}

\newcommand{\pseudoseg}{\textsc{pseudoseg}}
\newcommand{\algebraic}{\textsc{algebraic}}

\newcommand{\R}{\mathbb{R}} % real field
\newcommand{\OO}{\widetilde{O}}  
\newcommand{\Ohat}{O^*} 
\newcommand{\OOO}{\Ohat}
\newcommand{\cD}{\mathcal{D}} % a data structure
\newcommand{\eps}{\varepsilon} % epsilon
\renewcommand{\deg}{\textsf{deg}} %{d} % degree of the polynomials
\newcommand{\bad}{bad} % the bad set

\newcommand{\ignore}[1]{}

\title{Semialgebraic Range Stabbing, Ray Shooting, and Intersection Counting in the Plane}

\author{Timothy M. Chan}{University of Illinois Urbana-Champaign, Urbana, USA}{tmc@illinois.edu}{https://orcid.org/0000-0002-8093-0675}{Work supported by NSF Grant CCF-2224271.}
\author{Pingan Cheng}{Aarhus University, Aarhus, Denmark}{pingancheng@cs.au.dk}{https://orcid.org/0000-0002-8131-847X}{Work supported by DFF (Det Frie Forskningsr\aa d) of Danish Council for Independent Research under grant ID DFF–7014–00404 and STIBOFONDENs IT-rejsestipendier til ph.d.-studerende during the author's visit to UIUC.}
\author{Da Wei Zheng}{University of Illinois Urbana-Champaign, Urbana, USA}{dwzheng2@illinois.edu}{https://orcid.org/0000-0002-0844-9457}{}
\authorrunning{T. Chan, P. Cheng, and D. Zheng} %TODO mandatory. First: Use abbreviated first/middle names. Second (only in severe cases): Use first author plus 'et al.'
\Copyright{Timothy M. Chan, Pingan Cheng, and Da Wei Zheng} %TODO mandatory, please use full first names. LIPIcs license is "CC-BY";  http://creativecommons.org/licenses/by/3.0/

\titlerunning{Semialgebraic Range Stabbing, Ray Shooting, \& Intersection Counting in the Plane}

\relatedversion{}

 %TODO mandatory: Please choose ACM 2012 classifications from https://dl.acm.org/ccs/ccs_flat.cfm 

 %TODO mandatory; please add comma-separated list of keywords

\funding{}

\category{} %optional, e.g. invited paper

\begin{document}

\maketitle

\medskip

%documentclass[../main.tex]{subfiles}

\begin{abstract}

%We show new upper bounds for semialgebraic range stabbing.

\emph{Polynomial partitioning} techniques have recently led to improved geometric data structures for a variety of fundamental problems related to semialgebraic range searching and intersection searching in 3D and higher dimensions (e.g., see [Agarwal, Aronov, Ezra, and Zahl, SoCG 2019; Ezra and Sharir, SoCG 2021; Agarwal, Aronov, Ezra, Katz, and Sharir, SoCG 2022]).  They have also led to improved algorithms for \emph{offline} versions of semialgebraic range searching in 2D, via \emph{lens-cutting} [Sharir and Zahl (2017)].  In this paper, we show that these techniques can yield new data structures for a number of other 2D problems even for \emph{online} queries:

\begin{enumerate}
\item \emph{Semialgebraic range stabbing.}  We present a data structure for $n$ semialgebraic ranges in 2D of constant description complexity with $O(n^{3/2+\varepsilon})$ preprocessing time and space, so that we can count the number of ranges containing a query point in $O(n^{1/4+\varepsilon})$ time, for an arbitrarily small constant $\varepsilon>0$. (The query time bound is likely close to tight for this space bound.)

\item \emph{Ray shooting amid algebraic arcs.}  We present a data structure for $n$ algebraic arcs in 2D of constant description complexity with $O(n^{3/2+\varepsilon})$ preprocessing time and space, so that we can find the first arc hit by a query (straight-line) ray in $O(n^{1/4+\varepsilon})$ time. (The query bound is again likely close to tight for this space bound, and they improve a result by Ezra and Sharir with near $n^{3/2}$ space and near $\sqrt{n}$ query time.)

\item \emph{Intersection counting amid algebraic arcs.} We present a data structure for $n$ algebraic arcs in 2D of constant description complexity with $O(n^{3/2+\varepsilon})$ preprocessing time and space, so that we can count the number of intersection points with a query algebraic arc of constant description complexity in $O(n^{1/2+\varepsilon})$ time. In particular, this implies an $O(n^{3/2+\varepsilon})$-time algorithm for counting intersections between two sets of $n$ algebraic arcs in 2D. (This generalizes a classical $O(n^{3/2+\varepsilon})$-time algorithm for circular arcs by Agarwal and Sharir from SoCG 1991.)
\end{enumerate}

\end{abstract}

\section{Introduction}

The polynomial partitioning technique~\cite{gk15,g15} 
has led to 
a series of breakthroughs of many long-standing classic problems 
in computational geometry e.g., range searching~\cite{ams13,mp15,aaez21}, 
range stabbing~\cite{aaez21}, 
intersection searching~\cite{es22a,es22b,aaeks22}, etc, 
and simplification and generalization of 
many existing techniques and tools~\cite{kms12}. 
Comparing to rather simple geometric objects 
formed by halfspaces or hyperplanes 
that have been studied extensively 
in the early days of computational geometry, 
polynomial partitioning enables us to attain similar results for semialgebraic sets 
(a set obtained by union, intersection, and complement 
from a set of a collection of polynomial inequalities
where the number of polynomials,
the number of indeterminates, and
the degree of polynomials are constant).
Almost all of these breakthrough results are for 
problems in three or higher dimensions.
We complement these breakthroughs with some new results for 
fundamental problems involving algebraic curves in the plane.

\subsection{Problems studied and related results}

We consider the following three problems in this paper.

\subparagraph*{Semialgebraic range stabbing.}
In this problem we are given a collection of semialgebraic sets of constant complexity in $\R^2$ as the input,
and we want to preprocess them in a data structure so that we can quickly count or report
the inputs intersected or ``stabbed'' by a query point (this is called  a ``range stabbing query'', also known as a ``point enclosure query'').
Generalizing counting, we can also consider the \emph{semigroup model},
where every semialgebraic set is given a value in a semigroup, and we wish to apply the semigroup operation on the values of all sets stabbed.
Semialgebraic range stabbing and its ``dual'' problem,  semialgebraic range searching,
are among the most classical problems in computational geometry.
The two problems are relatively well-understood 
for linear ranges after a decade of study by pioneers in the fields 
in late 80s and early 90s.
We refer the readers to a survey of this topic~\cite{a16}.
The tools and results developed for the problems 
have also become textbook results~\cite{4m08}.

However, when considering general polynomial inequalities,
the problem is more difficult.
Before the invention of polynomial partitioning~\cite{gk15,g15},
there was a lack of suitable tools and
few tight results were known~\cite{am94}.
It was only very recently~\cite{ams13,mp15,aaez21}, 
that via polynomial partitioning, 
efficient data structures for the two problems were found
for data structures with small (near-linear) space,
and data structures with very fast (polylogarithmic) query time.
By interpolating the two extreme solutions,
we obtain space-time trade-offs.
However, somewhat mysteriously, 
even if the extreme cases are almost tight,
it is unknown whether the trade-off is close to optimal.
For example, even for the planar annulus stabbing,
there is a clear gap between the current upper bound\footnote{In this paper, 
we use the notation $\OOO(\cdot)$ or $\Omega^*(\cdot)$ to hide factors of $n^{\varepsilon}$ where $\eps>0$ is an arbitrary small constant.  We use the notation $\OO(\cdot)$ or $\widetilde\Omega(\cdot)$ to hide
factors polylogarithmic in $n$.
%and functions of $\varepsilon$
%for small positive constants $\varepsilon$ that we can make arbitrarily small.
} of $S(n)=\OOO(n^2/Q(n)^{3/2})$
and the lower bound of $S(n)=\Omega^*(n^{3/2}/Q(n)^{3/4})$~\cite{ac21}
or $S(n)=\widetilde\Omega(n^2/Q(n)^2)$~\cite{a13}, where $S(n)$ and $Q(n)$ denote space and query time respectively.

We mention that sometimes it is possible to
solve certain range searching problems involving algebraic arcs more efficiently.
For example, Agarwal and Sharir~\cite{as05} gave improved algorithms for counting 
containment pairs between points and circular disks in $\R^2$,
%intersections as efficiently
%as point-line segment intersections in the offline setting,
which can be viewed as an off-line version of either circular range searching or range stabbing.
To get this improvement, they used a key technique 
known as ``lens cutting'' to cut planar curves into pseudo-segments.
This allows us to use some of the classic tools
developed for linear objects which are usually 
more efficient than their polynomial counterparts.
However, to define the dual of pseudo-line or pseudo-segment arrangements,
we need to know all the input and query objects in advance;
that is the main reason why previous applications
are restricted to offline settings.
There were attempts to 
apply this technique to online problems~\cite{g09}
but to our knowledge they have not been generally successful.

\subparagraph*{Ray shooting amid algebraic arcs.}
We consider the problem of ray shooting
 where we are given a collection of algebraic arcs (of constant complexity) in $\R^2$ as the input, 
and we want to build a structure such that for any query (straight-line) ray,
we can find the first arc intersecting it or assert that no such arc exists.
Ray shooting is another classic problem in computational geometry with many applications in other fields such as computer graphics and robotics.
Early study of ray shooting mostly centered around special cases,
e.g., the input consists of line segments~\cite{a92,am93}, 
circular arcs~\cite{ako91}, or disjoint arcs~\cite{ako91}.
Specifically, for ray shooting queries amid line segments,
it is possible to obtain a trade-off of $S(n)=\OOO(n^2/Q(n)^2)$,
which has been conjectured to be close to be optimal.
For general algebraic curve inputs,
it is possible to build an $\OOO(n^2)$ space data structure 
with $O(\log n)$ query time in time $\OOO(n^2)$~\cite{k03}.
Combining the standard linear-space 
$O(n^{1-1/\beta})$-query time structure,
we can interpolate and get a space-time trade-off curve of
$S(n)=\OOO(n^2/Q(n)^{\beta/(\beta-1)})$, where
$\beta$ is the number of parameters needed to define
any polynomial in the semialgebraic sets
(for bivariate polynomials of degree $\deg$,
we have $\beta \le \binom{\deg+2}{2}-1$, but in general $\beta$ is often much smaller).
Very recently, Ezra and Sharir~\cite{es22a} showed how
 to answer ray shooting queries for algebraic curves
of constant complexity in $\R^2$ with $\OOO(n^{3/2})$ space
and $\OOO(n^{1/2})$ query time, where the exponent is independent of $\beta$.
Note that this gives better $\OOO(n^{3/2})$-space
data structures for all $\beta>3$.

\subparagraph*{Intersection counting amid algebraic arcs.}
Finally, we consider intersection counting amid algebraic arcs in $\R^2$---more precisely, computing the sum of the number of intersections between pairs of algebraic arcs.
We show new results for both online and offline versions of the problem.
For the online version where we want to build data structures
to count intersections with a query object, it is known that when the query object is a line segment,
a structure of space-time trade-off of 
$S(n)=\OOO(n^2/Q(n)^{3/2})$
(resp.\ $S(n)=\OOO(n^2/Q(n)^{\beta/(\beta-1)})$)
is possible for circular arcs (resp. general algebraic arcs)~\cite{k03} in the plane.
To the best of our knowledge,
the more general problem of algebraic arc-arc intersection counting has not been studied 
for offline intersection counting
where we are given a collection of algebraic arcs
and want to count the number of intersections points.
When the input consists of circular arcs,
there is an $\OOO(n^{3/2})$-time algorithm 
for the problem~\cite{AgarwalPS93}.
For more general arcs, it is unclear if any subquadratic
algorithm with exponent independent of $\beta$ exists.

\subsection{New results}

We present improved results for these three basic  problems 
in 2D computational geometry.

\begin{figure}
    \centering
    \includegraphics[]{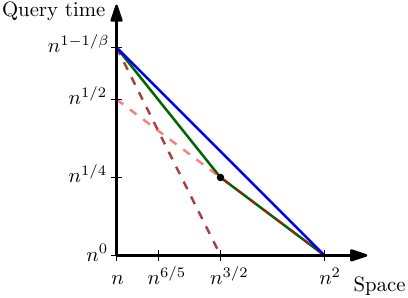}
    \caption{The blue line shows the prior known trade-off curve for semialgebraic range stabbing, and the green curve shows the improved trade-off curve we obtain. 
    The dotted red lines show the lower bounds of Afshani \cite{a13} for simplex stabbing and Afshani and Cheng~\cite{ac21} for semialgebraic range stabbing, both of which apply.}
    \label{fig:tradeoff}
\end{figure}

\subparagraph*{Semialgebraic range stabbing.}
We give a data structure with
$\OOO(n^{3/2})$ preprocessing time and space
and $\OOO(n^{1/4})$ query time
for semialgebraic range stabbing in $\R^2$. 
(This holds for counting as well as the semigroup model; for reporting, we add an $O(k)$ term to the query time where $k$ is the output size.)
Interestingly, the exponents here are independent of
the number $\beta$ of parameters  needed to define the algebraic curves 
(similar phenomena have recently been seen for certain problems in 3 and higher dimensions~\cite{es22a,es22b}).
The result matches known offline results
(namely, a batch of queries with $n^{5/4}$ ranges on $n$ points take
$\OOO(n^{3/2})$ total time~\cite{AgarwalPS93,aaez21}).
By interpolating with existing results,
we also automatically get an improved trade-off curve
for the (online) problem.
In particular, when the query time is at most $n^{1/4}$,
we obtain a space-time trade-off of 
$S(n)Q(n)^2=\OOO(n^2)$.
See \Cref{fig:tradeoff} for an illustration of the trade-off curve.
Note that it almost matches the curve for
simplex range stabbing, and is thus likely almost optimal, in this regime.
Prior to our result, online data structures matching this trade-off are known only when the query time
is very small (polylogarithmic), by constructing the entire $O(n^2)$-sized arrangement of the  ranges.

\subparagraph*{Ray shooting amid algebraic curves.}
We present a data structure with $\OOO(n^{3/2})$ preprocessing time and space 
that is able to answer ray shooting queries in time $\OOO(n^{1/4})$,
improving the $\OOO(n^{1/2})$ query time  
of the previous best structure
by Ezra and Sharir~\cite{es22a}.\footnote{To be fair, Ezra and Sharir's paper mainly focused on 3D versions of the ray shooting problem, and their 2D data structure was just one ingredient needed.  However, they did consider their 2D result to be ``of independent interest''.}
This again allows us to show a space-time trade-off 
that matches the one for 
ray shooting amid line segments 
when the query time satisfies $Q(n)=\OOO(n^{1/4})$.
Again, prior to our result, 
the trade-offs for the two problems only roughly match
for polylogarithmic query time structures.

\subparagraph*{Intersection counting amid algebraic arcs.}
For the online version where we need to preprocess a collection of algebraic arcs 
so that we can count the intersection among them and a query algebraic arc,
we give a structure with $\OOO(n^{3/2})$ preprocessing time and space
and $\OOO(n^{1/2})$ query time.
Prior to our work, such structure is only known for when
the query is a line segment instead of an algebraic arc.
A straightfoward application of our online result
immediately gives an $\OOO(n^{3/2})$ time algorithm
for the offline problem of counting the intersections of $n$ algebraic arcs.
This generalizes a known result for circular arcs~\cite{AgarwalPS93},
as well as the line segment-arc intersection detection
result by Ezra and Sharir~\cite{es22b}.

An interesting combinatorial consequence of our algorithm is that intersection graphs of algebraic arcs in $\R^2$ admit \emph{biclique covers}~\cite{AgarwalAAS94,FederM95} of size $\OOO(n^{3/2})$; it is again surprising that the exponent here is independent of $\beta$.  Biclique covers have many applications to algorithmic problems about geometric intersection graphs.

\subparagraph*{Concurrent work.}
%Concurrent to our work, 
In an independent work,
Agarwal, Ezra, and Sharir~\cite{aes24} showed that offline semigroup range searching with $m$ semialgebraic ranges with $\beta$ degrees of freedom 
and $n$ points in $\R^2$
can be solved in time $\OOO(m^{\frac{2\beta}{5\beta-4}}n^{\frac{5\beta-6}{5\beta-4}} + m^{2/3}n^{2/3}+m+n)$.
Furthermore, they show how to compute a biclique partition of the incidence graph between the semialgebraic sets and the points. 
We remark that the trade-offs we get for \emph{online} semialgebraic range stabbing \ARXIV{(in \Cref{ap:tradeoff}) }directly imply both of their results.
%also implies a biclique partition of the incident graph.

\section{Semialgebraic Range Stabbing}\label{sec:stab}
Let $\Gamma$ be a set of $n$ semialgebraic ranges in $\R^2$ where the boundary of each range consists of $O(1)$ algebraic arcs of degree at most $\deg=O(1)$.
In this section,
we present data structures to count \ARXIV{or report }the ranges stabbed by a query point. 
%In \Cref{sec:stabbing_reporting}, we present our algorithms for reporting the ranges and for queries in the semigroup model.

\subsection{Preliminaries}
We begin by reviewing known techniques for handling stabbing problems. 
One approach is by using $(1/r)$-cuttings~\cite{ClarksonS89,cf90,c93}.
\begin{lemma}[$(1/r)$-Cutting Lemma] \label{lem:cutting}
Given $n$ $x$-monotone algebraic arcs of constant degree in $\R^2$ and a parameter $r\le n$, there exists a decomposition of the plane into $O(r^2)$ disjoint pseudo-trapezoid cells such that each cell is crossed by at most $n/r$ arcs.

The cells, the list of arcs crossing each cell, and the number of arcs completely below each cell,
can all be computed in $O(nr)$ time.
\end{lemma}

Another method is based on the simplicial partition theorem:
\begin{theorem}[Matou\v{s}ek's Partition Theorem~\cite{m92a}] \label{lem:partition}
Let $P$ be a set of $n$ points in $\R^d$. Then for any $r\le n$, we can partition $P$ into $r$ disjoint simplicial cells such that each cell contains $O(n/r)$ points and any hyperplane crosses at most $O(r^{1-1/d})$ cells. This partition can be computed in $O(n)$ when $r$ is a constant.
\end{theorem}
To get a linear-space data structure, one approach is to lift the input curves to a halfspace in dimension 
$L={\deg+2\choose 2}-1$,
%$\beta$ (recall $\beta$ is the number of parameters needed to specify the curves) as in \cite{am94} 
and then apply the partition theorem recursively with constant $r$ in the dual to get a data structure with $\OOO(n)$ space and preprocessing time and 
$\OOO(n^{1-1/L})$ query time
(the extra $\OOO(1)$ factors can be lowered or removed \cite{c12}).
The query time bound can be improved
to $\OOO(n^{1-1/\beta})$ (recall that $\beta$ is the number of parameters needed to specify the curves),
by using an analog of the partition theorem for semialgebraic ranges for $\beta\le 4$~\cite{am94,k03},
or by using the polynomial partitioning method~\cite{ams13,mp15}.

Still %it is possible to do 
better results are possible if we have more guarantees about the behavior of the arcs of $\Gamma$.  The key case we will consider is
when the arcs form a set of \emph{pseudo-lines} ($x$-monotone curves, from $x=-\infty$ to $x=\infty$, that pairwise intersect at most once), or 
\emph{pseudo-segments} ($x$-monotone arcs that pairwise intersect at most once).

It turns out that the general case can be reduced to
the pseudo-line or pseudo-segment case by a technique known as \emph{lens cutting}, first proposed by Tamaki and Tokuyama~\cite{TamakiT98} and further developed by others~\cite{AgarwalNPPSS04,Chan03,MarcusT06}.
We use the
following theorem of Sharir and Zahl~\cite{sz17}
for cutting algebraic curves into pseudo-segments. 
The algorithmic version is due to Agarwal, Aronov, Ezra, and Zahl~\cite{aaez21}.
\begin{theorem}[Lens cutting for algebraic curves]
\label{thm:lens-cut}
    Given a collection $\Gamma$ of $n$ curves generated from constant-degree bivariate polynomials
	where no pair of polynomials shares a common factor,
	we can cut $\Gamma$ into a collection of $\OOO(n^{3/2})$ subarcs
	such that each pair of arcs intersect at most once.\footnote{If a pair of arcs $\gamma_1$ and $\gamma_2$ intersect more than once, the part of the two arcs between two consecutive intersections is sometimes referred to as a \emph{lens}.  Hence, the problem of cutting curves into pseudo-segments is also called \emph{lens cutting}.}
    Furthermore, this can be computed in $\OOO(n^{3/2})$ time.
\end{theorem}

Sharir and Zahl's theorem is striking in that it gives the first subquadratic bound for general algebraic arcs (previous results were for pseudo-parabolas~\cite{MarcusT06} or graphs of univariate polynomials~\cite{Chan03}), and at the same time, achieves an exponent ($3/2$) completely independent of the degree of the arcs!
%
%Therefore, in the next subsection, 
By lens cutting, we can thus turn our attention to solving the stabbing problem for ranges defined by pseudo-lines or pseudo-segments.

\subsection{Counting pseudo-lines below a query point} \label{sec:ps-counting}

We present our main new data structure for pseudo-lines below:

\begin{theorem} \label{thm:ps-counting}
    Given a set $\Gamma$ of $n$  pseudo-lines%
    \ARXIV{
    \footnote{We only assume that primitive operations such as deciding whether a point is above a pseudo-line, and computing the intersection of two pseudo-lines, can be done in constant time.} 
    }
    in $\R^2$,
    there is a data structure for counting the number of pseudo-lines below a query point  with $\OOO(n)$ preprocessing time and space and $\OOO(\sqrt{n})$ query time.
\end{theorem}

One approach to proving this theorem is via ``spanning trees with low crossing number''~\cite{Welzl88,w92}.
Chazelle and Welzl \cite{ChazelleW89} actually showed that such spanning trees can yield  range searching data structures in a general bounded-VC-dimension setting; our problem fits their framework, and so we can
immediately obtain a data structure with $O(\poly(n))$ preprocessing time, $\OO(n)$ space, and $\OO(\sqrt{n})$ query time for our problem.
We won't discuss this any further as it will be subsumed by our new approach which has much better preprocessing time and also has the advantage of supporting multi-level data structures (needed in our applications later).

Instead, our approach is based on dualizing Matou\v sek's partition theorem.
Recall that a standard way to solve the problem of counting lines (not pseudo-lines) below a query point
is to apply point/line duality to reduce the problem to counting points above a query line (i.e., halfplane range searching), which can then be solved using Matou\v{s}ek's partition tree.
Agarwal and Sharir \cite{as05} showed that there exists a similar duality between points and pseudo-lines.
However, this duality transform is only applicable when we know all the query points in advance---we can't dualize a new query point without potentially changing the entire transform.
Nonetheless, we have found a way to
overcome this issue.

We say that a point $p$ \emph{crosses} $S$ if there is at least one pseudo-line in $S$ above $p$, and at least one pseudo-line in $S$ below $p$. 
It turns out the right way to reformulate Matou\v sek's partition theorem in the dual is the following, whose proof requires several delicate steps:

\begin{theorem} \label{thm:pl_partition}
Given a set $\Gamma$ of $n$ pseudo-lines in $\R^2$ and a parameter $r\le n$, there exists a partition of $\Gamma$ into $r$ disjoint subsets $\Gamma_1,\ldots, \Gamma_r$ each of size $\Theta(n/r)$, such that any point crosses at most $O(\sqrt{r})$ of these subsets.
Furthermore, this partition can be computed in $O(nr^{O(1)})$ time.
\end{theorem}

\begin{proof}%[Proof of \Cref{lem:pl_partition}]
We start with a version of Matou\v sek's partition theorem, which follows directly from his original proof~\cite{m92a}
(see also the generalization in \cite[Lemma 5.2]{am94}):

\begin{description}
\item[(I)] Given a set $P$ of $n$ points in $\R^2$, a set $Q$ of $t$ ``test'' pseudo-lines, and a parameter $r\le n$,
there exists a partition of $P$ into $r$ disjoint subsets $P_1,\ldots, P_r$ each of size $\Theta(n/r)$,
together with $r$ (pseudo-trapezoidal) cells $\Delta_1,\ldots,\Delta_r$ with $P_i\subset \Delta_i$, such that any pseudo-line in $Q$
intersects at most $O(\sqrt{r} + \log t)$ of the cells.  
\end{description}

Next, we state a version that does not involve the cells $\Delta_i$ (this will be crucial, as it would be difficult to dualize $\Delta_i$).  Say that
a pseudo-line $\gamma$ \emph{crosses} a point set $P$ if $\gamma$ is above at least one point of $P$ and below
at least one point of $P$.

\begin{description}
\item[(II)] Given a set $P$ of $n$ points in $\R^2$, a set $Q$ of $t$ ``test'' pseudo-lines, and a parameter $r\le n$,
there exists a partition of $P$ into $r$ disjoint subsets $P_1,\ldots, P_r$ each of size $\Theta(n/r)$,
such that any pseudo-line in $Q$
crosses at most $O(\sqrt{r} + \log t)$ of the subsets.  
\end{description}

Observe that (II) follows from (I), since $P_i\subset \Delta_i$ implies that any pseudo-line crossing $P_i$ must intersect~$\Delta_i$.

Now, we apply the point/pseudo-line duality transform by Agarwal and Sharir~\cite{as05}, which turns (II) into the following statement:

\begin{description}
\item[(III)] Given a set $\Gamma$ of $n$ pseudo-lines, a set $M$ of $t$ ``test'' points, and a parameter $r\le n$,
there exists a partition of $\Gamma$ into $r$ disjoint subsets $\Gamma_1,\ldots, \Gamma_r$ each of size $\Theta(n/r)$,
such that any point in $M$
crosses at most $O(\sqrt{r} + \log t)$ of the subsets.  
\end{description}

The construction time for the partition in (I), and thus (II), is naively bounded by $O(n(rt)^{O(1)})$
from Matou\v sek's work~\cite{m92a}.
Unfortunately, the construction time for (III) is larger, since Agarwal and Sharir's duality transform
requires $O((nt)^{O(1)})$ time to compute~\cite{as05} (they obtained faster algorithms only under certain restricted settings).

We describe a way to speed up the construction for (III).
Say that two pseudo-lines $\gamma$ and $\gamma'$ are \emph{equivalent} with respect to $M$ if the subset of points of $M$ below $\gamma$
is identical to the subset of points of $M$ below $\gamma'$.
The problem of computing the equivalence classes of pseudo-lines with respect to a point set 
has luckily already been addressed in a paper by Chan~\cite[Sections 2.1--2.2]{Chan21a} (which studied a seemingly unrelated
problem: selection in totally monotone matrices).
Chan observed that the number of equivalence classes is $O(t^2)$ (this follows either by using
Agarwal and Sharir's duality transform to reduce to counting cells in the dual arrangement, or by 
direct VC dimension arguments), and he presented a simple deterministic $\OO(n+t^3)$-time algorithm
and a simple randomized $\OO(n+t^2)$-time algorithm
(by incrementally adding points of $M$ one by one and splitting equivalence classes using
dynamic data structures for lower/upper envelopes of pseudo-lines).

Afterwards, we can replace each pseudo-line with a representative member of its equivalence class.
As a result, we get a multi-set $\Gamma'$ of size $n$ that has only $O(t^2)$ distinct pseudo-lines.
We apply Agarwal and Sharir's duality transform to $\Gamma'$ and $M$, which now takes only $t^{O(1)}$ time.
We obtain a partition satisfying (III) for $\Gamma'$, which is automatically a partition satisfying (III) for $\Gamma$ by the definition of equivalence.
The overall construction time is $O(n (rt)^{O(1)})$.

Finally, we construct an appropriate (small) test set $M$ to establish our theorem.
The idea is similar in spirit to Matou\v sek's ``test set lemma''~\cite{m92a} (though his lemma is not directly applicable here).
We first compute a $(1/(cr))$-cutting of $\Gamma$ with $O(r^2)$ cells in $O(nr^{O(1)})$ time for a sufficiently large constant $c$;
each cell is a pseudo-trapezoid, with two vertical sides and the upper/lower sides being sub-segments
of the given pseudo-lines.  
We just define $M$ to be the set of all vertices of  these cells, with $t=|M|=O(r^2)$,
and construct the partition in (III) for this test set $M$ in $O(n(rt)^{O(1)})=O(nr^{O(1)})$ time.

Consider an arbitrary point $q\in\R^2$.  Let $\Delta$ be the pseudo-trapezoid cell containing $q$, 
with top-left vertex $v_{TL}$, bottom-left vertex $v_{BL}$, top-right vertex $v_{TR}$, and
bottom-right vertex $v_{BR}$.
Consider one subset $\Gamma_i$.
Suppose that none of $v_{TL},v_{BL},v_{TR},v_{BR}$ crosses $\Gamma_i$.  We prove that $q$ cannot cross $\Gamma_i$:

\begin{itemize}
\item Case 1: all pseudo-lines in $\Gamma_i$ are between $v_{TL}$ and $v_{BL}$.
Then all pseudo-lines in $\Gamma_i$ intersect $\Delta$, and so $|\Gamma_i|\le n/(cr)$, which is 
a contradiction if we choose $c$ large enough (compared to the hidden constant in the $\Theta(n/r)$ bound).
\item Case 2: all pseudo-lines in $\Gamma_i$ are above $v_{TL}$.
If all pseudo-lines in $\Gamma_i$ are also below $v_{TR}$, then all pseudo-lines in $\Gamma_i$ intersect $\Delta$ and
we again get a contradiction as in Case~1.
Thus, we may assume that all pseudo-lines in $\Gamma_i$ are above  both $v_{TL}$ and $v_{TR}$.
But then all pseudo-lines in $\Gamma_i$ are completely above $\Delta$ (since no pseudo-line
can intersect the upper side twice), and 
so $q$ cannot cross $\Gamma_i$.
\item Case 3: all pseudo-lines in $\Gamma_i$ are below $v_{BL}$.  Similar to Case 2.
\end{itemize}

We conclude that the subsets $\Gamma_i$ crossed by $q$ must be crossed by one of
the test points $v_{TL},v_{BL},v_{TR},v_{BR}$,
and so there are at most $O(4\cdot (\sqrt{r}+ \log t))=O(\sqrt{r})$ such subsets.
\end{proof}

\begin{proof}[Proof of \Cref{thm:ps-counting}]
%the inputs can be viewed as pseudo-lines, and thus~\Cref{lem:pl_partition} applies.

%\emph{Data Structure}
We construct a partition for $\Gamma$ by \Cref{thm:pl_partition}.
For each subset $\Gamma_i$, 
we store its upper and lower envelopes (which, for pseudo-lines, have $O(n)$ complexity and can be constructed in $O(n\log n)$ time, e.g., by a variant of Graham's scan~\cite{4m08}).
We recursively build the data structure for each $\Gamma_i$.

%\emph{Query Algorithm}
Given a query point $q$,
we examine each subset $\Gamma_i$.
If $q$ is below the lower envelope of $\Gamma_i$ (which we can check by binary search in $\OO(1)$ time),
we ignore $\Gamma_i$.
If $q$ is above the upper envelope of $\Gamma_i$,
we add $|\Gamma_i|$ to the current count.
Otherwise, $q$ crosses $\Gamma_i$, and we recursively query $\Gamma_i$.

Let $P(n)$ and $Q(n)$ be the preprocessing time and query time of the data structure (space is bounded by the preprocessing time).
They satisfy the following recurrence relations:
\begin{eqnarray*}
    P(n) &=& O(r)\cdot P(n/r) + \OO(r^{O(1)}n)\\
    Q(n) &=& O(\sqrt{r})\cdot Q(n/r)+\OO(r).
\end{eqnarray*}
Setting $r$ to be a large enough constant,
we obtain $P(n)=\OOO(n)$ and $Q(n)=\OOO(\sqrt{n})$.
\end{proof}

By using a segment tree~\cite{4m08}, we can easily extend \Cref{thm:ps-counting} to handle pseudo-segments:

\begin{corollary} \label{cor:ps-counting}
    Given a set $\Gamma$ of $n$ pseudo-segments in $\R^2$,
    there is a data structure for counting the number of pseudo-segments below a query point  with $\OOO(n)$  preprocessing time and space and $\OOO(\sqrt{n})$ query time.
\end{corollary}

\subsection{Semialgebraic range stabbing counting}

%We begin with by defining some subproblems.

\begin{definition}
For an integer $j\ge 0$ we say that $\Gamma$ is a set of $(j, \algebraic)$-ranges if each range is 
%defined by an $x$-interval $[a,b]$ as well as 
being bounded above/below by $j$ different $x$-monotone algebraic curves and at most two vertical sides.
We say that $\Gamma$ is a set of $(j, \pseudoseg)$-ranges if furthermore,
these $j$ curves are pseudo-segments.
%$\Gamma$ is a set of $(i, \algebraic)$-ranges and the $x$-monotone subarcs defining the ranges form a set of pseudo-segments. 
\end{definition}
For any set of $n$ semialgebraic ranges, we can decompose each range vertically with $O(1)$ cuts so that we get a set of $O(n)$ many $(2, \algebraic)$-ranges.
For counting, it suffices to look at a set of $(1, \algebraic)$-ranges with only lower bounding arcs, since we can use subtraction%
\ARXIV{
\footnote{In the more general semigroup model, subtraction would not be allowed. See \Cref{sec:stabbing_reporting} for how to directly handle $(2, \algebraic)$ ranges.}
}
to express a range bounded from above and below by the difference of two ranges bounded from below. %by above.

We reduce $(1,\algebraic)$-range stabbing to
$(1,\pseudoseg)$-range stabbing by lens cutting.
Naively replacing $n$ by $\OOO(n^{3/2})$ would yield 
terrible space and query bounds.
Past applications of lens cutting~\cite{AgarwalNPPSS04,as02,Chan03} first derived intersection-sensitive results for pseudo-segments, and noticed that the lens-cutting operation does not increase the number of intersections.  Below, we describe a direct reduction bypassing intersection-sensitive bounds:
%, which will be helpful later when considering multi-level data structures:

\begin{theorem}\label{thm:stabbing_counting}
    There is a data structure for range stabbing counting on $n$ semialgebraic ranges of constant complexity in $\R^2$ with $\Ohat(n^{3/2})$ preprocessing time and space and $\Ohat(n^{1/4})$ query time.
\end{theorem}
\begin{proof}
Let $\Gamma$ be a set of $n$
lower arcs (extended with upward vertical rays at their endpoints) of the  input ranges.
Compute a set of $\mu$ cut points that turn
their lower arcs into pseudo-segments.
We have $\mu=\OOO(n^{3/2})$ by \Cref{thm:lens-cut}.
Compute a $(1/r)$-cutting $\Xi$ of $\Gamma$ with $O(r^2)$ cells by \Cref{lem:cutting}.
Add extra vertical cuts to ensure that each cell contains at most $\mu/r^2$ cut points; the
number of cells remains $O(r^2)$.
For each cell $\Delta\in\Xi$, let $\Gamma_\Delta$ be the arcs in $\Gamma$ intersecting $\Delta$ (we know $|\Gamma_\Delta|\le n/r$); build the data structure
in \Cref{cor:ps-counting} for the $O(n/r+\mu/r^2)$
pseudo-segments along the arcs in $\Gamma_\Delta$ inside $\Delta$.  Let $c_\Delta$ be the number of arcs in $\Gamma$ completely below $\Delta$.
The preprocessing time/space is $\OOO(nr + r^2\cdot (n/r + \mu/r^2)) = \OOO(nr + \mu)$.

To answer a query for a point $q$, we find the cell $\Delta$ containing $q$ in $\OO(1)$ time by point location~\cite{4m08}, query the data structure for the pseudo-segments inside $\Delta$, and add $c_\Delta$ to the current count.
The query time $\OOO(1 + \sqrt{n/r+\mu/r^2})$.
Setting $r=\lceil \mu/n\rceil$ gives
preprocessing time/space $\OOO(n+\mu)$
and query time $\OOO(n/\sqrt{\mu})$.
\end{proof}

By standard techniques, we can use this to obtain improvement on the entire trade-off curve between space and query time.%
\ARXIV{
For completeness, we include a proof in \Cref{ap:tradeoff}.
}%
\SoCGV{
(See the full paper.)
}%
For semialgebraic stabbing reporting, we can no longer use subtraction and need to consider $(2, \algebraic)$-ranges.
Instead we can use multi-level data structures. 
%To do so, we can recursively use $(1/r)$-cuttings where $r=O(1)$.
This procedure is not straightforward, as we do not have a smooth tradeoff curve.%
\ARXIV{
Details will be given in the next subsection.
%See \Cref{sec:stabbing_reporting} for the details.
}%
\SoCGV{
Details are given in the full paper.
}

\subsection{Semialgebraic stabbing reporting}
\label{sec:stabbing_reporting}

In this subsection, we show how to adapt our data structure for  semialgebraic stabbing counting to solve the reporting version of the problem.  As noted, it suffices to consider $(2,\algebraic)$-ranges.  The idea is to use multi-level data structuring techniques.

First, we adapt \Cref{thm:ps-counting}/\Cref{cor:ps-counting} to handle $(2,\pseudoseg)$-ranges.
This part is straightforward.

\begin{lemma} \label{lem:2pseudoseg}
    There is a data structure for the semialgebraic stabbing reporting problem for $n$ $(2,\pseudoseg)$-ranges with $\OOO(n)$  preprocessing time and space, and $\OOO(\sqrt{n}+k)$ query time, where $k$ is the number of reported ranges.
\end{lemma}
\begin{proof}
By segment trees~\cite{4m08}, we may assume that all the pseudo-segments are pseudo-lines.
Let $\Gamma^-$ denote the set of lower arcs of the ranges in $\Gamma$.
We construct a partition for $\Gamma^-$ by \Cref{thm:pl_partition}.
For each subset $\Gamma_i$, 
we store its upper and lower envelopes.
We recursively build the data structure for the ranges corresponding to each $\Gamma_i$,
and also build a data structure for the $(1,\pseudoseg)$-ranges with upper arcs corresponding to the lower arcs in $\Gamma_i$.

%\emph{Query Algorithm}
Given a query point $q$,
we examine each subset $\Gamma_i$.
If $q$ is below the lower envelope of $\Gamma_i$ (which we can check by binary search in $\OO(1)$ time),
we ignore $\Gamma_i$.
If $q$ is above the upper envelope of $\Gamma_i$,
we query the $(1,\pseudoseg)$-ranges corresponding to $\Gamma_i$.
Otherwise, $q$ crosses $\Gamma_i$, and we recursively query $\Gamma_i$.

We can do the same to reduce the case of $(1,\pseudoseg)$-ranges to the trivial case of 
$(0,\pseudoseg)$-ranges.

For $j\in\{0,1,2\}$, let $P_j(n)$ be the preprocessing time of the data structure, and
let $Q_j(n)$ be the query time (ignoring the $O(k)$ term for the cost of reporting).
They satisfy the following recurrence relations:
\begin{eqnarray*}
    P_j(n) & =& O(r)\cdot P_j(n/r) + O(r)\cdot P_{j-1}(n) + \OO(r^{O(1)}n)
\\
    Q_j(n) &=& O(\sqrt{r})\cdot Q_j(n/r)+ O(r)\cdot Q_{j-1}(n) + \OO(r).
\end{eqnarray*}
Setting $r$ to be an arbitrarily large constant,
we obtain $P_j(n)=\OOO(n)$ and $Q_j(n)=\OOO(\sqrt{n})$.
\end{proof}

Now, we adapt \Cref{thm:stabbing_counting}.
This part is more delicate and requires a careful analysis: we don't know in general how to support multi-leveling in the structure from \Cref{thm:stabbing_counting} without losing efficiency, but in this application, we exploit the fact that we can map the cut points on the upper arcs to the lower arcs.

\begin{theorem}\label{thm:stabbing_reporting}
    There is a data structure for semialgebraic stabbing reporting problem for $n$ ranges of constant complexity in $\R^2$  with $\Ohat(n^{3/2})$  preprocessing time and space and  $\Ohat(n^{1/4}+k)$ query time, where $k$ is the number of reported ranges.
\end{theorem}
\begin{proof}
Let $t$ be a fixed parameter.  
Let $\Gamma$ be a set of $n$ $(2,\algebraic)$-ranges, together with a set of $\mu$ cut points that turn
both the upper algebraic arcs into pseudo-segments and the lower algebraic arcs into pseudo-segments.
In what follows, we assume that for each range, its upper arc and lower arc are cut at the same $x$-values, so that
the pseudo-segments on the two arcs are matched up in their $x$-ranges (this requires initially doubling the number of cut points).
Initially, $\mu=\OOO(n^{3/2})$ by \Cref{thm:lens-cut}.
Let $\Gamma^-$ denote the set of lower arcs (extended with upward vertical rays at the endpoints) of the ranges in $\Gamma$.
Compute a $(1/r)$-cutting $\Xi^-$ of $\Gamma^-$ with $O(r^2)$ cells for a sufficiently large constant $r$.
Add extra vertical cuts to ensure that each cell contains at most $\mu/r^2$ cut points on the lower arcs; the
number of cells remains $O(r^2)$.
For each cell $\Delta\in\Xi^+$, let $\Gamma_\Delta$ be the ranges in $\Gamma$ whose lower arcs
intersect $\Delta$ (we know $|\Gamma_\Delta|\le n/r$), and let $\Gamma'_\Delta$ be the ranges in $\Gamma$ whose lower arcs are completely below $\Delta$.  For each $\Delta$,
we recursively build a data structure for $\Gamma_\Delta$,
and also build a data structure for the $(1,\algebraic)$-ranges
with upper arcs corresponding to the ranges in $\Gamma'_\Delta$.
To answer a query for a point $q$, we find the cell $\Delta$ containing $q$, and recursively query $\Gamma_\Delta$, and also query the $(1,\algebraic)$-ranges corresponding to $\Gamma'_\Delta$.

When $n = \Theta(t^2)$, we stop the recursion, and solve the problem directly by dividing into
$\lceil \mu/n\rceil$ slabs with $O(n)$ cut points each, viewing the input in each slab as  $O(n)$ $(2,\pseudoseg)$-ranges, and applying our data structure for pseudo-segments (\Cref{lem:2pseudoseg}) with $\OOO(\lceil \mu/n\rceil \cdot n) = \OOO(n+\mu)$ total space and preprocessing time  
and  $\OOO(\sqrt{n})=\OOO(t)$ query time.  

We can do the same to reduce the case of $(1,\algebraic)$-ranges to the trivial case of $(0,\algebraic)$-ranges.

For $j\in\{0,1,2\}$, let $P_j(n,\mu)$ be the  preprocessing time of our data structure for $(j,\algebraic)$-ranges,
and let $Q_j(n,\mu)$ be the query time (ignoring the $+k$ term for the cost of reporting).  We obtain the recurrences:
\[
P_{j}(n,\mu) \:=\: 
\begin{cases}
O(r^2) \cdot P_{j} (n/r, \mu/r^2) + O(r^2) \cdot S_{j-1} (n, \mu) + \OOO(r^2(n + \mu)) & \text{ if } n =\Omega(t^2) \\
\OOO(n+\mu) & \text{ if } n= \Theta(t^2), \\
\end{cases}
\]
\[
Q_{j}(n,\mu) \:=\: 
\begin{cases}
 Q_{j} (n/r, \mu/r^2) + Q_{j-1} (n, \mu) + O(r^2) & \text{ if } n =\Omega(t^2) \\
\OOO(t) & \text{ if } n=\Theta(t^2), \\
\end{cases}
\]
For $n=\Omega(t^2)$,
the recurrences solve to $P_j(n,\mu) = \OOO(n^2/t^2 + \mu)$ (noting that $n\le O(n^2/t^2)$) and $Q_j(n,\mu)=\OOO(t)$, by choosing $r$ to be an arbitrarily large constant.
Thus, the overall data structure has $\OOO(n^2/t^2 + n^{3/2})$ space and preprocessing time and $\OOO(t)$ query time.
Finally, we set $t=n^{1/4}$.
\end{proof}

The same approach works in the semigroup model (without the $+k$ term).
\section{Ray Shooting Amid Curves}\label{ap:ray:shoot:long}

As an application of our range stabbing result, we describe an algorithm for ray shooting amid curves. 
Let $\Gamma$ be a collection of $n$ algebraic arcs of degree at most $\deg=O(1)$. By breaking each arc into a constant number of subarcs, we may assume each arc is $x$-monotone and either convex or concave.
W.l.o.g., we assume all arcs are convex. 
Furthermore, we assume all query rays are non-vertical. 

By parametric search~\cite{am93,Megiddo83} or randomized search~\cite{Chan99}, it suffices to focus on the problem where we wish to detect if a query line segment intersects any of the input arcs. 
We will present a data structure that can more generally count the number of intersections between a query line segment $\overline{pq}$ and the input arcs.
With subtraction, it suffices to count the number of intersections between the arcs and two query rightward rays (a ray emanating from $p$ and a ray emanating from $q$).
To handle this, we begin by building a segment tree structure~\cite{4m08,ChazelleEGS94} on the input arcs.
It suffices to focus on solving the intersection counting problem in these two cases in a slab $\sigma$ of the segment tree: 
(i) ``long-short intersections'' when the query rays span the entire slab but the input arcs may not, and 
(ii) ``short-long intersections'' when 
the input arcs span the entire slab but the query rays may not.
All other intersections may be handled by recursion.

\subsection{Counting long-short intersections}\label{sec:line-arc}
\begin{figure}
    \centering
    \includegraphics[height=0.2\textheight]{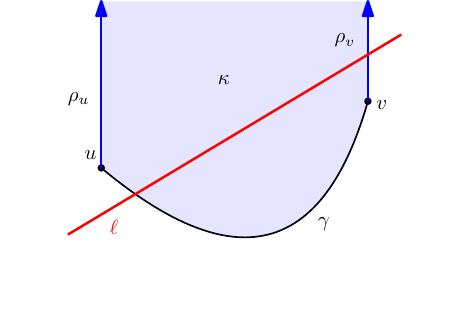}
    \vspace*{-5ex}
    \caption{The line $\ell$ intersecting $\kappa$ intersects $\rho_v$ and $\gamma$.}
    \label{fig:line-arc}
\end{figure}

To handle case (i), we may treat the query ray as a line $\ell$.
For simplicity we will assume that $\ell$ is non-vertical and is not tangent to any of the arcs of $\Gamma$. 
For each (convex) arc $\gamma\in\Gamma$,
we define $\kappa(\gamma)$ to be the (convex) region
bounded by $\gamma$ and the upward vertical rays $\rho_u$ and $\rho_v$
emanating from the two endpoints $u$ and $v$ of $\gamma$.
It is easy to see that the number of intersections between $\ell$ and $\gamma$ 
is exactly  the number of intersections between $\ell$ and $\kappa(\gamma)$,
minus the number of intersections between $\ell$ and $\rho_u$ and between $\ell$ and $\rho_v$.  (See \Cref{fig:line-arc}.)

Intersections between $\ell$ and the vertical rays $\rho_u$ and $\rho_v$ can be easily counted in $O(n^{1/4})$ query time using $O(n^{3/2})$ space and preprocessing by known data structures for halfplane range counting in $\R^2$ \cite{m92a,cz22}.

Thus, it suffices to count intersections between $\ell$ and the regions $\{\kappa(\gamma): \gamma\in \Gamma\}$.
Here, we use duality.
A line $\ell$ intersects the convex region $\kappa(\gamma)$ that is unbounded from above if and only if the dual point $\ell^*$ lies in the dual convex region $\kappa^*(\gamma)$ that is unbounded from below. Note that $\kappa^*(\gamma)$ is a semialgebraic range with two rays and an arc $\gamma^*$ which is a subarc of the \emph{dual curve} of $\gamma$, a curve with degree $O(\deg^2)$ \cite{brieskorn2012plane}.
Thus,
%as all such dual regions are semialgebraic ranges (of slightly higher but still constant degree), 
we can apply \Cref{thm:stabbing_counting} to count the number of such ranges stabbed by $\ell^*$.
The overall query time is $\OOO(n^{1/4})$ with $\OOO(n^{3/2})$ space and preprocessing time.

\subsection{Counting short-long intersections} % between a ray and long arcs within a slab}

In case (ii), the input arcs span  the entire $x$-range $[x_1,x_2]$ of the slab $\sigma$. 
We may assume that each input arc $\gamma$ has endpoints $u$ and $v$ with $u_x = x_1$ and $x_v=x_2$,
and by clipping, the query ray becomes a query line segment $\overline{pq}$ with $q_x = x_2$.
Let $\ell$ be the line extension of $\overline{pq}$.
%Let $m$ and $b$ be the slope and $y$-intercept of $\ell$.

We first describe how to count the number of arcs $\gamma$ that intersect the query segment $\overline{pq}$ \emph{exactly once}.
This happens iff

\begin{figure}
    \centering
    \includegraphics[width=0.3\textwidth, page=1]{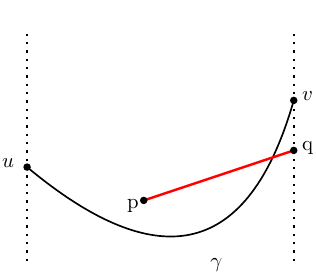}
    \includegraphics[width=0.3\textwidth, page=3]{ray_arc.pdf}
    \includegraphics[width=0.3\textwidth, page=2]{ray_arc.pdf}
    \caption{Three different ways a ray $\rho$ can intersect an arc $\gamma$. The left depicts Case A, the middle depicts Case B, and the right depicts Case C.}
    \label{fig:ray_arc}
\end{figure}

\begin{itemize}
    \item Case A: $q_y < v_y$ and $p$ is above $\gamma$, or
    \item Case B: $q_y > v_y$ and $p$ is below $\gamma$.  (See Figure~\ref{fig:ray_arc}.)
\end{itemize}

Consider Case A.
As the first condition is a ``one-dimensional'' constraint and the second corresponds to range stabbing for the regions $\kappa(\gamma)$,
we can just use a 2-level data structure,
where the primary structure is a 1D range tree~\cite{4m08}
and the secondary structure is
the structure from \Cref{thm:stabbing_counting}.
The data structure has $\OOO(n^{1/4})$ query time
and $\OOO(n^{3/2})$ space and preprocessing time.
Case B is the same.

The more difficult subproblem is how to count the number of arcs $\gamma$ that intersect the query segment $\overline{pq}$ \emph{exactly twice}.
We observe that this happens iff the following four conditions are simultaneously true (this is similar to an observation by Ezra and Sharir \cite{es22b}):

\begin{itemize}
%    \item[(a)] Once, if $q_y < v_y$ and $p$ is above $\gamma$, 
%    \item[(b)] Once, if $q_y > v_y$ and $p$ is below $\gamma$, 
    \item Case C: (i)~$q_y < v_y$, and (ii)~$p$ is below $\gamma$, and (iii)~$\ell$ intersects $\kappa(\gamma)$, and (iv)~the slope of $\gamma$ at $p_x$ is less than the slope of $\ell$.
    (See Figure~\ref{fig:ray_arc}.)
\end{itemize}
%\end{observation}

%Case (c) is considerably more difficult.
Condition (i) %that $q_y < v_y$
can be handled with the approach as in Case A or B\@.
This leaves three conditions remaining that we claim all correspond to semialgebraic stabbing problems.
Condition (ii) corresponds to a stabbing problem in the \emph{original space}.
%the point $p$ is below the curve $\gamma$.
Condition (iii) 
%that $\ell$ intersects the region $\kappa(\gamma)$ 
corresponds to a stabbing problem in the \emph{dual space}, as it is equivalent to the condition that the dual point $\ell^*$ lies in %beneath
the dual region $\kappa^*(\gamma)$,  as we have already discussed in \Cref{sec:line-arc}
%is a semialgebraic range stabbing problem involving the dual arc $\gamma^*$ 
(technically the semialgebraic range can be vertically decomposed into regions above the dual arc $\gamma^*$ and parts above certain lines, but the line portion is easily handled by known multi-level data structures).
The final condition (iv)
corresponds to a stabbing problem in the \emph{tangent space}.
If we let $y=f(x)$ %$f(x,y) = 0$ 
be the equation defining the arc $\gamma$ for an algebraic function~$f$, and let $y=mx+b$ be the  equation defining the line $\ell$,
this problem in the tangent space 
is equivalent to the condition that the point $(p_x, m)$ lies above the curve:
\[  \partial\gamma\ = \{\left(x,\tfrac{df}{dx}(x)\right): x_1\le x\le x_2\}.\]
This is an algebraic curve of degree at most $O(\deg^2)$ \cite{brieskorn2012plane}. 

Thus the whole problem is somewhat similar to a $(3, \algebraic)$-range stabbing problem (albeit over 3 different spaces). 
However, as discussed in \Cref{sec:stabbing_reporting},
we can't directly construct a multi-level data structure for semialgebraic range stabbing because the trade-off curve is not smooth.
Fortunately, even though the three conditions involve different query points in the original space, the dual space, and the tangent space, all three conditions are related to a single curve $\gamma$.
This means that we can apply the lens cutting algorithm of \Cref{thm:lens-cut} in the three respective spaces, and map all cut points back to the curve $\gamma$ in the original space.
This can be summed up in the following lemma.

\begin{lemma}
For any set of $n$ algebraic arcs $\Gamma$ of constant complexity in $\R^2$, it can be cut into a collection $\Gamma'$ of $\Ohat(n^{3/2})$ subarcs, in $\Ohat(n^{3/2})$ time,
so that for any two subarcs $\gamma_1, \gamma_2 \in \Gamma'$:

\begin{enumerate}
    \item[(a)] the curves $\gamma_1$ and $\gamma_2$ intersects in at most once.
    \item[(b)] the derivative curves $\partial \gamma_1$ and $\partial \gamma_2$ intersect at most once.
    \item[(c)] the dual curves $\gamma^*_1$ and $\gamma^*_2$ intersect at most once.
\end{enumerate}
\end{lemma}

Thus, by an analysis very similar to that in \Cref{sec:stabbing_reporting} (but with 3 levels instead of 2), we can build a data structure for Case~C with $\Ohat(n^{1/4})$ query time and $\Ohat(n^{3/2})$ preprocessing time and space. 
%This concludes our result.

\begin{theorem}
Given $n$ algebraic arcs of constant complexity in $\R^2$,
there is a data structure with $\OOO(n^{3/2})$ preprocessing time and space that can count intersections with a query line segment in $\OOO(n^{1/4})$ time. 
Consequently, there is a data structure for ray shooting amid $n$ algebraic arcs of constant complexity in $\R^2$ with
the same bound.
%$\Ohat(n^{3/2})$ preprocessing time and space and  $\Ohat(n^{1/4})$ query time.
\end{theorem}

\section{Intersection Counting Amid Algebraic Arcs}
\label{sec:intersect}

Let $\Gamma$ be a set of $n$ algebraic arcs in $\R^2$ of degree at most $\deg = O(1)$.
In this section we present algorithms and data structures for counting the number of intersections between the arcs of $\Gamma$.
By the ``number of intersections'', we always mean the number of intersection points (possibly with multiplicities if we have degeneracies/tangencies), and not the number of intersecting pairs.%
\ARXIV{
We assume that no two algebraic arcs lie in the same algebraic variety so that the number of intersections between two arcs is at most $\deg^2 = O(1)$.
}
W.l.o.g., we assume that the arcs are $x$-monotone.

\subsection{A first approach}
To better appreciate our final algorithm, we  first sketch a slower but still subquadratic algorithm for the offline problem of counting  intersections among $n$ algebraic arcs in $\R^2$, by using the lens cutting routine of \Cref{thm:lens-cut} as a black box as we did in earlier sections.  Let $r$ be a parameter to be chosen later.

\begin{enumerate}
    \item Compute a $(1/r)$-cutting $\Xi$ of $\Gamma$ into $O(r^2)$ disjoint cells each intersecting $n/r$ arcs.
    \item Compute a set $P$ of $\mu=\Ohat(n^{3/2})$ points to chop the arcs of $\Gamma$ into  pseudo-segments by \Cref{thm:lens-cut}, in $\OOO(n^{3/2})$ time. 
    Refine the cutting $\Xi$ by adding vertical line segments so that each cell of the cutting contains at most $\mu/r^2$ points of $P$; the number of cells remain $O(r^2)$. The number of pseudo-segments in each cell is bounded by the number of arcs intersecting each cell and the number of points of $P$ in the cell, which is $m=O(n/r+\mu/r^2)$.
    %and thus is at most $m = \Ohat(n^{1/2})$. 
    %Furthermore, we've added $O(n)$ vertical lines so the number of cells of $\Xi^*$ is still $O(n)$. 
    \item In each cell, use an $\Ohat(m^{4/3})$-time algorithm to count the number of intersections between pseudo-segments (algorithms are known \cite{m92a,cz22} for counting intersections between line segments in near $m^{4/3}$ time, and they can be adapted to pseudo-segments as well, but we will not elaborate as we will present a better algorithm shortly).
\end{enumerate}

\noindent
The total run time is thus $\OOO(r^2\cdot (n/r + \mu/r^2)^{4/3})$.
Choosing $r=\lceil\mu/n\rceil$ yields a time bound of
$\OOO(n+n^{2/3}\mu^{2/3})=\OOO(n^{5/3})$.

\ARXIV{
(We note that the above method works also for the case of pseudo-parabolas that are not necessarily algebraic arcs.  Marcus and Tardos \cite{MarcusT06} proved that $\mu=O(n^{3/2}\log n)$ cuts suffices.  However, the best algorithm for construct such cut points~\cite{Har-PeledS01,Solan98} requires time $\OO(n\sqrt{\mu})=\OO(n^{7/4})$, and so the running time increases to $\OO(n^{7/4})$.)
}

We will next show how to improve the running time for algebraic arcs to $\Ohat(n^{3/2})$, by opening the black box of \Cref{thm:lens-cut} and directly modifying the algorithm for cutting algebraic arcs into pseudo-segments.
%to get an algorithm for counting intersections for algebraic arcs in $\Ohat(n^{3/2})$ time. 
Remarkably, this algorithm naturally extends to a data structure for%
\SoCGV{
intersection counting.
}%
\ARXIV{
counting the number of intersections between a query algebraic arc and a set of $n$ input algebraic arcs with $\Ohat(n^{3/2})$  preprocessing time and space and $\Ohat(n^{1/2})$ query time.
(In contrast, known methods for cutting pseudo-parabolas \cite{MarcusT06,Har-PeledS01,Solan98} require all pseudo-parabolas to be given offline and cannot be adapted into such a data structure.)
}

\subsection{Review of lens cutting}\label{sec:review:lens:cut}

\newcommand{\gammahat}{\widehat{\gamma}}
\newcommand{\Gammahat}{\widehat{\Gamma}}

As a preliminary, we  sketch the proof of Sharir and Zahl~\cite{sz17} for \Cref{thm:lens-cut}.
The key idea is to transform the 2D  lens cutting problem  into a 3D problem about eliminating depth cycles, which was already solved before by Aronov and Sharir~\cite{AronovS18} using polynomial partitioning techniques.
Let $\Gamma$ be a set of $n$ algebraic $x$-monotone plane arcs each of degree at most $\deg=O(1)$. 
For any arc $\gamma \in \Gamma$ of the form $\{(x,f(x)): x_1\le x\le x_2\}$ for some algebraic function $f$ and $x_1,x_2\in\R$, we \emph{lift} $\gamma$ to
a new arc in 3D:
\[ \gammahat = \{\left(x,f(x), \tfrac{df}{dx}(x)\right): x_1\le x\le x_2\}. \]
In other words, the $xy$-projection of $\gammahat$ is $\gamma$, and the $z$-coordinate corresponds to the slope of the curve.  The arc $\gammahat$ is algebraic, with 
 degree at most $\deg^2$ (see Lemma 2.5 of \cite{sz17} or Proposition 1 of \cite{EllenbergSZ16} for precise details).
%a function $f(x,y) = 0$ is $x$-monotone and has no point where $\partial_y f(x,y) = 0$ by introducing $O(n)$ extra cuts.
Let $\Gammahat$ denote the set of these arcs in $\R^3$. 

To eliminate depth cycles in $\R^3$, Aronov and Sharir~\cite{AronovS18} proceeded by computing a \emph{polynomial partition} of $\Gammahat$, i.e.,  a polynomial $P(x,y,z)$ whose zero set $Z(P) := \{(x,y,z)\in \R^3: P(x,y,z) = 0\}$ separates $\R^3$ into cells such that not too many arcs of $\Gammahat$ intersect each cell. 
This was proved to exist by Guth \cite{g15} and the construction was made algorithmic by Agarwal, Aronov, Ezra, and Zahl~\cite{aaez21}. The theorem applies to general varieties in any dimension, but we will present it specialized to curves in $\R^3$.

\begin{theorem}[Polynomial partitioning of curves in $\R^3$]
Let $\Gamma$ be a collection of $n$ algebraic arcs in $\R^3$ each of which has degree at most $\deg=O(1)$. Then for any $D\ge 1$ there is a non-zero polynomial $P$ of degree at most $D$ such that $\R^3 \setminus Z(P)$ contains $O(D^3)$ cells and each cell crosses at most $O(n/D^2)$ arcs of $\Gamma$. 

The polynomial $P$ and the semi-algebraic representation of every cell $\R^3 \setminus Z(P)$ can be constructed in $O(2^{\poly(D)})$ randomized expected time. Furthermore this representation of the cells has size $O(\poly(D))$, and given any algebraic arc, we can output the cells of $\R^3\setminus Z(P)$ that  it crosses (or that it lies completely within $Z(P)$) in $O(\poly(D))$ time. In particular, we can compute the set of arcs intersecting every cell of $\R^3 \setminus Z(P)$ in $O(n\poly(D))$ time.
\end{theorem}

We proceed next by cutting each curve of $\Gammahat$ at its intersection points with the zero set $Z(P)$ of a partitioning polynomial $P$ of degree $D$.  We further cut each curve at its intersection points with
another surface $Z_{\bad}$ which is the vertical cylinder passing through all points with vertical tangency at $Z(P)$
(this is also a zero set of a polynomial, of degree $O(D^2)$).
%formed by the zero set of 
%another polynomial of degree $O(D^2)$ where the gradient of $P$ is vertical.
For any point $z$, let $h(z)$ denote the number of times a vertical downward ray emanating from $z$ intersects $Z(P)$.  Then points $z$ in the same cell
of $\R^3 \setminus (Z(P) \cup Z_{\bad})$ have the same $h(z)$ value,
by our definition of $Z_{\bad}$.
%For any cell of $\Delta \in \R^3 \setminus (Z(P) \cup Z_{\bad})$,
%let $h(z)$ for a point $z\in \Delta$ denote the number of times a vertical downward ray emanating from $z$ intersects $Z(P)$. By the choice of $Z_{\bad}$, for any other $z'\in \Delta$, we must have $h(z) = h(z')$.

The key observation is that two curves $\gamma_1, \gamma_2 \in \Gamma$ in 2D intersect twice if and only if their corresponding curves $\gammahat_1, \gammahat_2 \in \Gammahat$ in 3D form a length-2 \emph{depth cycle} in the $z$-direction, i.e., there are four points $(x, y, z_1), (x',y',z_1') \in \gammahat_1$, and $(x,y,z_2), (x',y', z_2') \in \gammahat_2$ where 
$z_1>z_2$ and $z_1'<z_2'$,
or vice versa. 
(This is because in 2D, at two consecutive intersection points between $\gamma_1$ and $\gamma_2$, the slope of $\gamma_1$ is larger  than the slope of $\gamma_2$ at one point, and vice versa at the other point.)

Recall that we have cut the arcs at intersections with $Z(P)$ as well as $Z_{\bad}$.  Suppose a subarc of $\gammahat_1$ is contained in a cell $\Delta$ of $\R^3\setminus Z(P)$, and suppose a subarc of $\gammahat_2$ is not contained in the same cell $\Delta$.  We observe that the two subarcs cannot form a length-2 depth cycle.
This is because otherwise,
%Furthermore, because we cut the arcs at intersections with $Z_{\bad}$, if subarcs of $\gammahat_1$ and $\gammahat_2$ are in different cells of $\R^3\setminus Z(P)$ but form a depth cycle, then 
$h(z_1) > h(z_2)$ and $h(z_1') < h(z_2')$, or vice versa, which is a contradiction.
%

%Thus there will be no depth cycles between two (cut) arcs that are completely contained in two different cells of $\R^3 \setminus Z(P)$.

%The intuition for why this works is that the cells of $\R^3 \setminus (Z_{\bad} \cup Z(P))$ below a given point are ``well-ordered'', i.e. each pair of cells intersecting a vertical line is above or below the other line.

%To handle depth cycles within a single cell, recursion is used. 
Thus, it suffices to eliminate length-2 depth cycles for pairs of arcs contained in the same cell $\Delta$ of $\R^3\setminus Z(P)$; this can be handled by recursion in each cell.
Arcs contained in $Z(P)$ or $Z_{\bad}$ can be handled naively as there can only be $O(D^2)$ %algebraic curves defining 
such arcs. The run time and number of cuts satisfy a recurrence of the form $T(n) = O(D^3) \cdot T(n/D^2) + O(n\poly(D) + 2^{\poly(D)})$. By choosing $D$ to be a sufficiently large constant, this recurrence solves to $T(n) = \Ohat(n^{3/2})$, thereby proving \Cref{thm:lens-cut}.

\subsection{An improved data structure for counting intersections}

In this section we directly adapt the approach in Section~\ref{sec:review:lens:cut}  to design a new data structure for counting  intersections between a query algebraic arc $\gamma$ and a set of $n$ input algebraic arcs $\Gamma$ in $\R^2$.%
%For simplicity we assume no two input arcs are subarcs of the same algebraic curve. If they are not, we may perturb all arcs by a small random amount upwards.

First we consider an easier special case, where we are guaranteed that the query arc $\gamma$ intersects each curve of $\Gamma$ at most once.  This special case will be useful later.
We prove the following lemma 
\SoCGV{in the full paper, }%
%in \Cref{ap:lemds_once_proof} 
by standard reductions to semialgebraic range searching and range stabbing (the bounds below may not be tight, but will be good enough): 
\begin{lemma} \label{lem:ds_once}
Given a set $\Gamma$ of $n$ algebraic arcs of constant complexity in $\R^2$,
there is a data structure with $\OOO(n^{3/2})$ preprocessing time and space that can count intersections with a query algebraic arc $\gamma$ in $\OOO(\sqrt{n})$ time if the query arc is guaranteed to intersect each arc of $\Gamma$ at most once.
\end{lemma}
\ARXIV{%%%%%%%%%%%%%%%
\begin{proof}
We can process the arcs with a segment-tree like approach similar to what we did previously for ray shooting. 
As we recurse with our query arc $\gamma_{q}$, we it suffices to consider the following two types of intersections:
(i) long-short intersections, where the query arc spans the entire slab but the input arcs may not, and 
(ii) short-long intersections, where the input arcs span the entire slab but the query may not.
%intersections in slabs where our query arc is short and the input arcs are long within the slab.

For intersections of type (i), the problem reduces to counting the number of input arcs that have one end point above $\gamma_{q}$ and the other below $\gamma_{q}$.
This can be done using a two-level version of the data structure of Agarwal and Matou\v{s}ek for semialgebraic range searching \cite{am94}, with $\OO(n)$ preprocessing time/space and  $\OOO(\sqrt{n})$ time.

For intersections of type (ii), the problem reduces to counting the number of input arcs that lie above one endpoint of $\gamma_q$ and below the other endpoint of $\gamma_q$.
This problem can be solved with a two-level data structure for range stabbing, with $\OO(n^2)$ preprocessing time/space and $\OO(1)$ query time by using cutting trees.
Conveniently, it works out that by splitting
the input to $\sqrt{n}$ groups of size $\sqrt{n}$,
the preprocessing time/space reduces to
$\OO(\sqrt{n}\cdot (\sqrt{n})^2)=\OO(n^{3/2})$,
while the query time increases to $\OO(\sqrt{n})$.
\end{proof}
}%%%%%%%%%%%%%%%%%%

We begin by considering $\Gammahat$, the lifted version of each arc in $\R^3$, and taking a polynomial partition $P$ of the curves of $\Gammahat$ with degree $D$, which we will choose to be a sufficiently large constant.
Let $\Gamma_{\bad}$ denote the set of bad arcs that, when lifted to $\R^3$, are contained in $Z(P)$ or $Z_{\bad}$ as defined in Section~\ref{sec:review:lens:cut}; there are at most $O(D^2)$ bad arcs.
For each cell $\Delta \in \R^3 \setminus Z(P)$, let $\Gammahat(\Delta)$ denote the set of all maximal subarcs of all $\gammahat \in \Gammahat \setminus \Gammahat_{\bad}$ that are contained in $\Delta$. 
Furthermore, let $\Gammahat'(\Delta)$ denote the set of subarcs of the arcs in $\Gammahat(\Delta)$ after cutting each arc at its intersections with $Z_{\bad}$. 
Let $\Gamma(\Delta)$ denote the $xy$-projections of the arcs of $\Gammahat(\Delta)$, and define  $\Gamma'(\Delta)$ similarly.
We recursively build our data structure for each $\Gamma(\Delta)$.
In addition, we preprocess each $\Gamma'(\Delta)$ in the data structure $\cD(\Delta)$ from \Cref{lem:ds_once}.
%for counting intersections between a query arc and the arcs of $\Gamma'(\Delta)$ if we know the query arc intersects each arc of $\Gamma'(\Delta)$ exactly once. 

\newcommand{\subarc}{\gamma'}
\newcommand{\subarchat}{\widehat{\gamma}'}

Given a query arc $\gamma_q$, we first cut $\gammahat_q$ into subarcs at intersections with $Z(P) \cup Z_{\bad}$; there are $O(D^2)$ such subarcs.  We cut $\gamma_q$ at the corresponding points.
%Aronov and Sharir's or Sharir and Zahl's analysis (see \Cref{sec:review:lens:cut}) shows that each subarc $\subarc_q$ of $\gamma_q$ can intersect each arc of $\Gamma'(\Delta)$ at most once if $\subarchat_q$ is not contained in $\Delta$ (i.e., if $\subarchat_q$ is contained in a different cell of $\R^3\setminus Z(P)$, or is inside $Z(P)$).
%This suggests the following query algorithm:
Our query algorithm is as follows:

\begin{enumerate}
    \item For each cell $\Delta \in \R^3 \setminus Z(P)$ crossed by $\gamma_q$:
    we count the intersections of $\gamma_q$ with $\Gamma(\Delta)$ by recursion.  There are $O(D)$ recursive calls, since $\gamma_q$ can cross $Z(P)$ at most $O(D)$ times.
    \item For each cell  $\Delta \in \R^3 \setminus Z(P)$ not crossed by $\gamma_q$, and for each subarc $\subarc_q$ of $\gamma_q$: we know that
    $\subarchat_q$ is not contained in $\Delta$.
    As we have observed in Section~\ref{sec:review:lens:cut},
    in this case, $\subarchat_q$ cannot form length-2 depth cycles with the subarcs in $\Gammahat'(\Delta)$, and thus
    $\subarc_q$ cannot form lenses with the subarcs in $\Gamma'(\Delta)$, i.e., $\subarchat_q$ can intersect each arc of $\Gamma'(\Delta)$ at most once.
    Using the data structure $\cD(\Delta)$ from \Cref{lem:ds_once}, we can count the intersections of $\subarc_q$ with $\Gamma'(\Delta)$.
    \item Finally, we naively count the intersections of $\gamma_q$ with $\Gamma_{\bad}$.  This takes $O(D^2)$ time.
\end{enumerate}

\noindent
This way, every intersection point along $\gamma_q$ is counted exactly once.

The query time satisfies the following recurrence:
\[ Q(n) \:=\: O(D) \cdot Q(n/D^2) + \OOO(D^{O(1)}\sqrt{n}),  \]
which solves to $Q(n)=\OOO(\sqrt{n})$
by choosing an arbitrarily large constant $D$.
The preprocessing time (and thus space) satisfies the following recurrence:
\[ P(n) \:=\: O(D^3)\cdot  P(n/D^2) +  \OOO(D^{O(1)}n^{3/2} + 2^{\poly(D)}), \]
which solves to $P(n) = \OOO(n^{3/2})$. 

\begin{theorem} \label{thm:acds}
Given $n$ algebraic arcs of constant complexity in $\R^2$,
there is a data structure with $\OOO(n^{3/2})$ preprocessing time and space that can count intersections with a query algebraic arc $\gamma$ in $\OOO(\sqrt{n})$ time.%
\SoCGV{
In particular, we can count the number of intersection points between two sets of $n$ algebraic arcs of constant complexity in $\R^2$ in $\OOO(n^{3/2})$ time.
}
\end{theorem}

\ARXIV{
This immediately implies an offline algorithm for counting the number of intersections among algebraic arcs\SoCGV{ in $\OOO(n^{3/2})$ time}.

\begin{corollary}
There is an algorithm that counts the number of intersection points between two sets of $n$ algebraic arcs of constant complexity in $\R^2$ in $\OOO(n^{3/2})$ time.
\end{corollary}
}

\section{Final Remarks}

We believe that the relative simplicity of our algorithm for counting arc intersections in Section~\ref{sec:intersect}
makes it a good example illustrating the power of the polynomial partitioning techniques
(and the 2D problem of counting arc intersections is in some sense even more basic than
the 3D problem of eliminating depth cycles or the 2D problem of cutting lenses).

%\textbf{Remark.} 
One application is an $\OOO(n^{3/2})$-time algorithm for verifying whether a set of algebraic arcs in $\R^2$ is a pseudo-line or pseudo-segment arrangement (or equivalently, detecting the existence of a lens).  We just check whether the total number of intersections is equal to
the number of odd-intersecting pairs, the latter of which can be computed using a variant of \Cref{lem:ds_once}.  

Our algorithm for arc intersection counting can be modified to give a \emph{biclique cover}~\cite{AgarwalAAS94,FederM95} (but not a biclique partition) of the intersection graph of $n$ algebraic arcs in $\R^2$ with size $\OOO(n^{3/2})$ in $\OOO(n^{3/2})$ time. Biclique covers are useful sparse representations of geometric intersection graphs, and our results imply many algorithmic results for algebraic arcs that use biclique covers (e.g., finding connected components in $\OOO(n^{3/2})$ time~\cite{Chan06}, 
finding single-source shortest paths in an intersection graph in 
$\OOO(n^{3/2})$ time,
finding maximum matching in a bipartite intersection graph in $\OOO(n^{3/2})$ time,\footnote{As the original paper by Feder and Motwani~\cite{FederM95} showed,
maximum matching in an unweighted bipartite graph with $n$ vertices and biclique cover size $s$ reduces to maximum flow with unit capacities in a graph of size $O(n+s)$, which can be solved in $\OOO(n+s)$ time by recent breakthrough results on maximum flow~\cite{ChenKLPGS22}.
}
or finding triangles in an intersection graph in subquadratic time~\cite{Chan23}).

Currently, we do not know how to modify our algorithm for counting arc intersections to 
count the number of intersecting pairs (except in the pseudo-parabola case).
This is also a weakness in some of the higher-dimensional results by Agarwal et al.~\cite{aaeks22}.

In Section~\ref{sec:stab}, we have shown how the lens cutting technique can be applied to obtain data structures
for online 2D semialgebraic range stabbing queries, but it remains open whether the
same is possible for online 2D semialgebraic range searching (when the query semialgebraic
ranges are not known in advance).

The lens cutting technique allows us to achieve the same bound for 2D semialgebraic range
stabbing as 2D simplex range stabbing for certain parts of the trade-off curve.
An intriguing question is whether the same is possible also in dimension 3 and higher,
via some generalization of lens cutting or some completely different technique.

\bibliography{reference}{}
\bibliographystyle{abbrvurl}
%\subfile{sections/appendices}

\appendix

\newcommand{\IGNORE}[1]{}

\IGNORE{%%%%%%%%%

\section{Proof of \Cref{lem:ds_once}}
\label{ap:lemds_once_proof}

\begin{proof}
We can process the arcs with a segment-tree like approach similar to what we did previously for ray shooting. 
As we recurse with our query arc $\gamma_{q}$, we it suffices to consider the following two types of intersections:
(i) long-short intersections, where the query arc spans the entire slab but the input arcs may not, and 
(ii) short-long intersections, where the input arcs span the entire slab but the query may not.
%intersections in slabs where our query arc is short and the input arcs are long within the slab.

For intersections of type (i), the problem reduces to counting the number of input arcs that have one end point above $\gamma_{q}$ and the other below $\gamma_{q}$.
This can be done using a two-level version of the data structure of Agarwal and Matou\v{s}ek for semialgebraic range searching \cite{am94}, with $\OO(n)$ preprocessing time/space and  $\OOO(\sqrt{n})$ time.

For intersections of type (ii), the problem reduces to counting the number of input arcs that lie above one endpoint of $\gamma_q$ and below the other endpoint of $\gamma_q$.
This problem can be solved with a two-level data structure for range stabbing, with $\OO(n^2)$ preprocessing time/space and $\OO(1)$ query time by using cutting trees.
Conveniently, it works out that by splitting
the input to $\sqrt{n}$ groups of size $\sqrt{n}$,
the preprocessing time/space reduces to
$\OO(\sqrt{n}\cdot (\sqrt{n})^2)=\OO(n^{3/2})$,
while the query time increases to $\OO(\sqrt{n})$.
\end{proof}
}%%%%%%%%%%%%%%%%%%%%

\section{Trade-Off Between Preprocessing and Query Time}\label{ap:tradeoff}

We note a trade-off version of our result on semialgebraic range stabbing counting:

\begin{theorem} \label{thm:stabbing_tradeoff}
    There is a data structure for the semialgebraic range stabbing counting problem on $n$ ranges of constant complexity in $\R^2$ 
    %defined by algebraic curves of degree at most $\deg=O(1)$ 
    with $\OOO(m)$  space and handles queries in time
    \[ Q(n) = \begin{cases}
    \OOO(n/\sqrt{m}) & \text{ if } n^{3/2} \le m \le n^{2} \\
    \OOO(n^{5/2-3/\beta}/m^{3/2-2/\beta}) & \text{ if } n \le m < n^{3/2}
    \end{cases}
    \]
    where $\beta$ is the number of parameters needed to specify a curve.
    Furthermore, the data structure can be constructed in randomized expected $\OOO(m)$ time.
\end{theorem}

\begin{proof}%[Proof of \Cref{thm:stabbing_tradeoff}]
Recall that for counting, it suffices to consider $(1,\algebraic)$-ranges.

First we consider the case when $n^{3/2} \le m\le n^{2}$.
The proof of Theorem~\ref{thm:stabbing_counting} already implies a trade-off with preprocessing time/space $\OOO(nr + n^{3/2})$
and query time $\OOO(1+\sqrt{n/r + n^{3/2}/r^2})$.
%This case we reduce to handling long arcs by constructing a cutting with parameter $r$.
%By adding additional vertical lines to cells we can ensure that the $O(r^2)$ cells intersects at most $\OOO(n/r + n^{3/2}/r^2)$ arcs, and the arcs that do intersect form a collection of $(1, \pseudoseg)$-ranges.
%Here, we can build the data structure of \Cref{thm:ps-counting} to get a data structure with $\OOO(nr))$ query time.
Setting $r=m/n$ gives the desired $\Ohat(m)$ preprocessing time/space  bound and $\Ohat(n/\sqrt{m})$ query time bound.

When $n \le m< n^{3/2}$, we instead map the input curves to points in the dual space $\R^\beta$.
For $\beta\le 4$, we can apply 
an analog of the partition theorem for semialgebraic ranges by Agarwal and Matou\v sek~\cite{am94,k03} to recursively 
split our instance into subproblems until each subproblem 
contains at most $b$ points for 
a parameter $b$.
For general constant $\beta$,
we instead apply 
%the multi-level partition theorem 
the polynomial partitioning method
of Matou\v{s}ek and Pat\'akov\'a~\cite{mp15}.
%for a parameter $D = O(1)$ to recursively 
%split our instance into subproblems until each subproblem 
%contains at most $b$ points for 
%a parameter $b$.
%$\ell = n^{3}/m^2$.
When a subproblem contains at most $b$ points, we switch back to the primal and build the data structure of \Cref{thm:stabbing_counting} with $\OOO(b^{3/2})$ preprocessing time and $\OOO(b^{1/4})$ query time. 
Agarwal et al.~\cite[Appendix~A.1]{aaeks22arxiv} provided details of the recursion and analysis based on the polynomial partitioning method (and even more generally in the multi-level setting), which we will not repeat here (since the only change is our new base case).
Let $P(n_v, s)$ and $Q(n_v,s)$ denote the expected preprocessing time and query time for a node of a partition tree with $n_v$ dual points that lie on the zero set of  dimension $s$.
With our new bounds for the base case, 
their recurrences \cite[Equations (24) and (25)]{aaeks22arxiv} (in the single-level case) are changed to the following:
\[ P(n_v, s)\ \le\  
\begin{cases}
O^*(n_v)   &\text{ if } s=1 \\
O(D)\cdot P(n_v/D, s) +  P(n_v, s-1) + O(n_v) &\text{ if } n_v\ge b \\
\OOO(b^{3/2}) &\text{ if } n_v < b
\end{cases}\]
\[ 
Q(n_v, s)\ \le\  
\begin{cases}
O^*(1)  &\text{ if } s=1 \\
O(D^{1-1/\beta})\cdot Q(n_v/D, s) +  Q(n_v, s-1) &\text{ if } n_v\ge b \\
\OOO(b^{1/4}) &\text{ if } n_v < b
\end{cases}\]
It can be verified that $P(n, \beta) = \OOO((n/b)\cdot b^{3/2})$ and $Q(n,\beta) = \OOO((n/b)^{1-1/\beta}\cdot b^{1/4})$.  Setting $b=(m/n)^2$ gives the desired bounds.
%The query time \cite[Equation 25]{aaeks22arxiv} satisfies the following recurrence:
\end{proof}
We remark that we can obtain a similar trade-off in the semi-group model, range reporting (with an additional $+k$ term), and ray shooting by using the appropriate multi-level versions of polynomial partitioning as in \cite[Appendix A.1]{aaeks22arxiv}.

\end{document}